\documentclass[11pt]{article}

\usepackage{graphicx}
\usepackage{float}
\usepackage{subcaption}
\usepackage{framed}

\usepackage{amsmath}
\usepackage{amssymb} 
\usepackage{amsfonts}
\usepackage{bm}
\DeclareSymbolFont{matha}{OML}{txmi}{m}{it}
\DeclareMathSymbol{\varv}{\mathord}{matha}{118}
\newcommand{\indep}{\perp \!\!\! \perp}
\newtheorem{theorem}{Theorem}

\usepackage{multirow}
\usepackage[flushleft]{threeparttable}
\usepackage{booktabs,caption}
\usepackage{hyperref}
\usepackage[round]{natbib}
\usepackage{url}

\newcommand{\blind}{1}

\begin{document}

\def\spacingset#1{\renewcommand{\baselinestretch}%
{#1}\small\normalsize} \spacingset{1}


\if1\blind
{
  \title{\bf A Randomization-Based Method for Evaluating Time-Varying Treatment Effects}
  \author{Sangjin Lee \\ 
  Department of Statistics, Seoul National University, Republic of Korea \\
  and \\
  Kwonsang Lee\thanks{Corresponding author: Kwonsang Lee (kwonsanglee@snu.ac.kr)}\hspace{.2cm}\\
  Department of Statistics, Seoul National University, Republic of Korea}
  \date{}
  \maketitle
} \fi

\if0\blind
{
  \bigskip
  \bigskip
  \bigskip
  \begin{center}
    {\LARGE\bf Evaluating Time-Specific Treatment Effects Using Randomization Inference}
\end{center}
  \medskip
} \fi

\bigskip


\begin{abstract}

Tests for paired censored outcomes have been extensively studied, with some justified in the context of randomization-based inference.
These tests are primarily designed to detect an overall treatment effect across the entire follow-up period, providing limited insight into when the effect manifests and how it changes over time.
In this article, we introduce new randomization-based tests for paired censored outcomes that enable both time-specific and long-term analysis of a treatment effect. The tests utilize time-specific scores, quantifying each individual's impact on sample survival at a fixed time, obtained via \textit{pseudo-observations}. 
Moreover, we develop corresponding sensitivity analysis methods to address potential unmeasured confounding in observational studies where randomization often lacks support.
To illustrate how our methods can provide a fuller analysis of a time-varying treatment effect, we apply them to a matched cohort study using data from the Korean Longitudinal Study of Aging (KLoSA), focusing on the effect of social engagement on survival.

\end{abstract}

\noindent%
\vfill


\newpage
\spacingset{1.9}


\section{Introduction}
\label{s:intro}

\subsection{Motivating Example}
\label{sec:motive}

\citet{kim2016} analyzed the Korean Longitudinal Study of Aging (KLoSA) cohort of individuals aged 45 years or older in 2006 to assess the effect of social engagement on survival.
Using a Cox proportional hazards model, they found that individuals with lower levels of social engagement had a higher risk of mortality, with an estimated hazard ratio of 1.84 for those in the lowest level of engagement compared to those in the highest.
While their approach relied on social behavioral trajectories between 2006 and 2012, we aim to assess the effect of social engagement at baseline in 2006 on subsequent survival in the cohort. 
In particular, our primary interest lies in identifying the actual duration of this effect, if any, beyond merely determining whether it existed. 
It should be acknowledged that the baseline measure of social engagement may not capture long-term behavioral patterns, nor can we ascertain that social behaviors were sustained over time.
Yet, the one-time measure provides a consistent reference point for adjusting baseline confounders \citep{lee2018powerful, zhang2024} and thereby enables what \citet{hernan2020} describe as an observational analog of intention-to-treat (ITT) analysis.

\begin{figure}[H]
\centering
\includegraphics[width=0.8\textwidth]{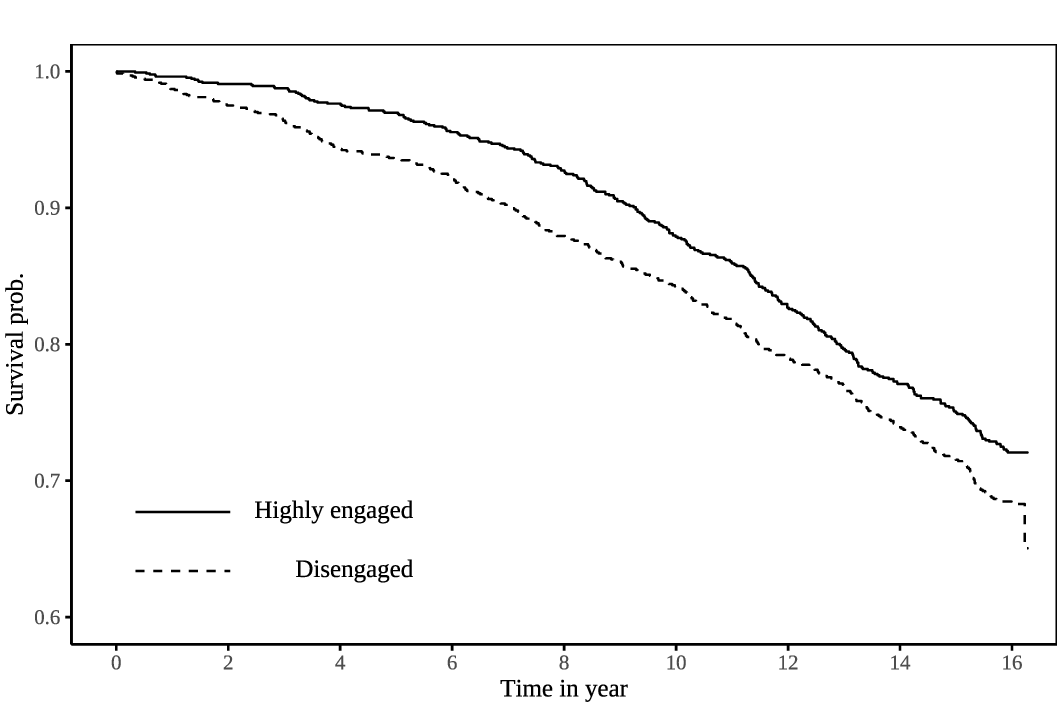}
\caption{Comparison of the Kaplan–Meier survival curves for highly engaged individuals and their matched disengaged controls in the 1,474 matched pairs from the KLoSA cohort}
\label{fig:km_klosa}
\end{figure}

We follow \citet{kim2016} in defining a binary treatment based on total social engagement scores (0-17), though we use only those from 2006. 
In addition, we define the outcome as time to death or right-censoring within the extended follow-up period through 2022.
Using the standard practice of matching to remove bias from measured confounding \citep{rosenbaum2002, Imbens2015}, we paired 1,474 treated individuals who were highly engaged (total score $>$ 6) with 1,474 control individuals who were disengaged (total score $\leq$ 3). 
Details on the matching variables and the resulting balance are provided in~\ref{sec:append_matching}. 
Figure~\ref{fig:km_klosa} shows the Kaplan–Meier survival curves for highly engaged individuals and their matched disengaged controls in the 1,474 matched pairs.
From the figure, the survival difference between the treated and control groups appears to be most pronounced between follow-up years 4 and 10 with the gap generally narrowing toward the end of follow-up. 
This may suggest that social engagement at baseline prolonged survival for a decade but had little or no effect on 13- or 14-year survival. However, this inspection alone does not provide valid evidence that a beneficial effect truly existed, nor does it clarify how the effect varied over time.  
To investigate this further, we will examine the matched pairs using our proposed method in Section~\ref{sec:application}.
It is also important to bear in mind that concerns about unmeasured confounding remain in our motivating example.
Because the KLoSA data are observational, there is no guarantee that two individuals in a matched pair had equal chances of being highly socially engaged in 2006, even though all measured confounders were adjusted for by matching. 
To address this, we will also investigate how unmeasured confounding might affect our inferences through sensitivity analyses.

\subsection{Our Contribution} \label{sec:appli}
\label{sec:contri}

Tests for paired censored outcomes have been extensively studied by \citet{obrien1987}, \citet{dabro1989}, \citet{akritas1992}, \citet{dallas2000}, \citet{murray2002}, among others, with some justified in the context of randomization-based inference. 
For example, the paired Prentice-Wilcoxon test of \citep{obrien1987} has been shown to be valid for testing Fisher’s sharp null hypothesis of no treatment effect at all \citep{rosenbaum2002}. 
Although focused on unpaired data, \citet{Li2023} provided a new justification of the classical log-rank test from a randomization-based perspective,
suggesting broader applicability of randomization-based reasoning to time-to-event analysis.

Regardless of the inferential framework under which they are developed, existing tests for paired censored outcomes generally share several limitations. 
While they may be effective in determining whether a treatment has an overall effect during follow-up, they often provide limited insight into how the effect manifests over time. This lack of time-specific information is particularly problematic in settings where the treatment effect is expected to vary—either gradually or abruptly—throughout the study period. In many scientific contexts, especially in clinical or behavioral research, treatment effects are seldom static; rather, they may intensify, diminish, or even reverse over time.
When such patterns remain unclear, so does our understanding of how the treatment works over time, precluding the identification of its onset or duration.
In addition, the statistical power of existing tests can vary considerably depending on the timing and shape of the underlying treatment effect \citep{woolson1992, sun1996, dallas2000, lee2003}, rendering the results sensitive to how the effect unfolds over time. For example, a test may perform well when the effect is strong and constant throughout follow-up, but suffer dramatically when the effect is delayed, short-lived, or non-monotonic. This sensitivity introduces a layer of uncertainty into the analysis: the same method that works well in one scenario may fail in another, simply because the underlying pattern of the effect has changed.
In an effort to address these limitations, complementary approaches have been proposed by focusing on comparing survival probabilities at a fixed time point \citep{klein2007, su2011}. However, these fixed-time methods lack a principled approach to formally identify the onset and duration of a treatment effect. Taken together, these observations highlight the need for a method that not only provides time-specific insight into a treatment effect but also maintains a valid inference and robust statistical power across a range of effect trajectories.

In this article, we introduce new randomization-based tests for paired censored outcomes that enable both time-specific and long-term analysis of treatment effects. A key idea of our approach is to use time-specific scores that quantify each individual’s contribution to sample survival at a fixed time point. These scores are derived from \textit{pseudo-observations}, which allow for the construction of interpretable, time-indexed test statistics. 
The two tests comprise: (i) a time-specific test to determine whether a treatment effect exists up to a given time $\tau$, and (ii) an overall test to determine whether any treatment effect exists over the entire follow-up period. Through simulation studies, we demonstrate that the overall test maintains strong power across a range of time-varying effect scenarios—including one with crossing survival curves—where many conventional methods struggle.
In addition to addressing time-varying treatment effects, our framework also allows for valid inference under deviations from strict randomization by incorporating sensitivity analysis methods based on Rosenbaum's assignment model for binary treatments \citep{rosenbaum1987, rosenbaum2002}. This approach builds upon the previous work by \citet{lu2018}.

\section{Notation and Review} \label{sec2}

\subsection{Notation: Randomization-Based Inference for Paired Censored Data} \label{sec:notation} 

We consider $I$ independent pairs, $i = 1,\dots,I$, of two units, $j = 1,2$, one treated indicated by $Z_{ij}=1$ and the other control indicated by $Z_{ij} = 0$, with $Z_{i1} + Z_{i2} = 1$ for each $i$. 
The pairs are matched for measured covariates, $\bm{x}_{i1} = \bm{x}_{i2}$, but may possibly differ in terms of an unmeasured covariate, $u_{i1} \neq u_{i2}$. 
Under the potential outcome framework \citep{neyman1923, rubin1974}, each unit $ij$ has two potential survival (event) times, $s_{1ij}$ and $s_{0ij}$, under treatment and control, respectively, so the realized survival time for $ij$ is given by $S_{ij} = Z_{ij}s_{1ij} + (1-Z_{ij})s_{0ij}$.
Also, we assume that each unit $ij$ has a deterministic right-censoring time, $C_{ij}$, invariant to the assignment of treatment $Z_{ij}$.
Note that ($S_{ij}$, $C_{ij}$) is not jointly observed for any $ij$ in practice. Instead, we observe the minimum $Y_{ij} = \min(S_{ij}, C_{ij})$, along with the indicator $\Delta_{ij} = I(S_{ij} \leq C_{ij})$ for whether the event occurs before censoring.
Write $\bm{Y} = (Y_{11}, \dots, Y_{I2})^T$, $\bm{\Delta} = (\Delta_{11}, \dots, \Delta_{I2})^T$ for the $2I$ dimensional vectors, with a similar notation for $\bm{u}$. 
For simplicity, the treated and control units in each pair $i$ are identified by the sign of $V_i = Z_{i1} - Z_{i2}$, with $V_i=1$ if $i1$ is treated and $V_i=-1$ if $i2$ is treated. 
Write $\bm{V} = (V_1,..., V_I)^T$ for the $I$ dimensional vector of treatment assignments for all $I$ pairs. 
Let $\mathcal{V}$ be the set containing the $2^I$ possible values of $\bm{V}$, so that $\bm{v}  \in \mathcal{V}$ if each $v_i$ is $1$ or $-1$. 
Moreover, write $\mathcal{F} = \{(s_{1ij},s_{0ij},C_{ij},x_{ij},u_{ij}),i = 1,\dots, I, j = 1,2\}$. 
Then, the sets $\mathcal{F}$ and $\mathcal{V}$ are fixed through the course of randomization-based inference \citep{fisher1935}.

In a paired experiment, the treatment is randomly assigned within each pair $i$, ensuring that $\Pr(\bm{V}=\bm{v} \mid \mathcal{F},\mathcal{V}) = 1/2^I$ for each $\bm{v} \in \mathcal{V}$. Thus, under the sharp null hypothesis,
\begin{equation*} 
H_0:s_{0ij} = s_{1ij} \quad \text{for any} \ ij, 
\end{equation*}
given a test statistic $T(\bm{V},\bm{Y},\bm{\Delta})$ and its observed value $t$, the randomization $p$-value is $\Pr(T \geq t \mid \mathcal{F},\mathcal{V}) = \lvert \{\bm{v} \in \mathcal{V}: T(\bm{v}, \bm{Y},\bm{\delta}) \geq t\} \rvert /2^I$ as ($\bm{Y}$, $\bm{\Delta}$) is fixed by conditioning on $\mathcal{F}$ and $\bm{V}$ is uniformly distributed on $\mathcal{V}$. 
To present a typical form of $T$, let $q_{ij}$ be a censored-data score assigned to unit $ij$ that depends only on $(\bm{Y}, \bm{\Delta})$. 
For instance, $q_{ij}$ might be the log-rank score \citep{mantel1966} defined by $\hat{H}(Y_{ij}) - \Delta_{ij}$, where $\hat{H}(a) = \sum_{k: Y_k \leq a} m_k/n_k$, $Y_k$ is an observed time among all $2I$ observations, $n_k$ is the number at risk at $Y_k$, and $m_k$ is the number of events that happened at $Y_k$. 
Alternatively, $q_{ij}$ might be the Prentice-Wilcoxon score \citep{kalbfleisch1980} defined as $1-\hat{J}(Y_{ij}) - \Delta_{ij}\hat{J}(Y_{ij})$, where $\hat{J}(a) = \prod_{k: Y_{k} \leq a} (n_k - m_k +1)/(n_k+1)$, with $Y_k$, $n_k$ and $m_k$ as defined. 
The test statistics proposed by 
\citet{obrien1987} and \citet{dallas2000} can then be written as the sum of treated-minus-control differences in scores, $T = \sum_{i=1}^{I} (q_{i1}-q_{i2})(Z_{i1}-Z_{i2}) = \sum_{i=1}^{I} d_iV_i$, where $d_i = q_{i1}-q_{i2}$.
In this statistic, the $q_{ij}$'s are fixed under $H_0$ by conditioning on $\mathcal{F}$ and the randomness comes solely from the assignment vector $\bm{V}$.
In particular, pairs with tied scores have $d_iV_i = 0$, regardless of the value of $V_i$, so they do not contribute any stochastic component to the statistic.

\subsection{Sensitivity Analysis in Observational Studies}
\label{sec:review_sensitivity_analysis}

In a observational study, failure to match on the unmeasured covariate $u_{ij}$ may lead to biased treatment assignments within pairs, $\Pr(\bm{V}=\bm{v} \mid \mathcal{F},\mathcal{V}) \neq 1/2^I$. A simple model for sensitivity to departures from random treatment assignments assumes that the distribution of $Z_{ij}$, given measured covariates $\bm{x}_{ij}$ and a hypothetical unmeasured covariate $u_{ij}$, follows a logistic model,
\begin{equation} 
\label{eq:logit_model}
\log \frac{\Pr(Z_{ij}=1|\bm{x}_{ij}, u_{ij})}{\Pr(Z_{ij}=0|\bm{x}_{ij}, u_{ij})} = \kappa(\bm{x}_{ij}) + \gamma u_{ij},
\end{equation} 
where $\kappa(\cdot)$ is an arbitrary function, $\gamma \geq 0$ (or  equivalently $\Gamma = \exp(\gamma) \geq 1$) is a sensitivity parameter, and $u_{ij} \in [0,1]$ is normalized to make $\gamma$ more interpretable \citep{rosenbaum1987, rosenbaum2002}. More precisely, \eqref{eq:logit_model} says that two units in pair $i$ will have odds of receiving treatment that differ by factor $1/\Gamma \leq \exp\{\gamma (u_{i1}-u_{i2})\} \leq \Gamma$ after matching on $\bm{x}_{ij}$. Write $w_i = u_{i1} - u_{i2}$ and $\bm{w} = (w_1,...w_I)^T$.
Then, it is easy to verify that \eqref{eq:logit_model} is the same as assuming that
\begin{equation} 
\label{eq:rosenbaum_model}
\Pr(\bm{V}=\bm{v} \mid \mathcal{F},\mathcal{V}) = \prod_{i=1}^{I} \frac{\exp(\frac{1}{2}\gamma v_iw_i)}{\exp(\frac{1}{2}\gamma w_i)+\exp(-\frac{1}{2}\gamma w_i)} = \frac{\exp(\frac{1}{2}\gamma \bm{v}^T\bm{w})}{\sum_{\bm{b} \in \mathcal{V}} \exp(\frac{1}{2}\gamma \bm{b}^T\bm{w})},
\end{equation} 
for each $\bm{v} \in \mathcal{V}$. In the absence of bias from $\bm{u}$, $\gamma=0$, \eqref{eq:rosenbaum_model} is the randomization distribution $\Pr(\bm{V}=\bm{v} \mid \mathcal{F},\mathcal{V}) = 1/2^I$, whereas if $\gamma>0$, then the probability $\Pr(\bm{V}=\bm{v} \mid \mathcal{F},\mathcal{V})$ is unknown because $\bm{w}$ is unmeasured. 
Although \eqref{eq:rosenbaum_model} cannot be used for direct adjustment of the $p$-value $\Pr(T \geq t \mid \mathcal{F},\mathcal{V})$, the \textit{worst-case $p$-value}, defined as the largest $p$-value over all possible allocations of $\bm{w} \in [-1,1]^I$, can be found as follows. Define $w_i^+=\text{sgn}(d_i)$, where $\text{sgn}(a)=1$ if $a$ is positive, -1 if $a$ is negative, and 0 if $a$ is zero. 
Write $\bm{w}^+ = (w_1^+,\dots,w_I^+)^T$. 
Let $T^+$ denote the distribution of $T = \sum_{i=1}^{I} d_iV_i$ when $\bm{w}$ equals $\bm{w}^+$. Then, under \eqref{eq:rosenbaum_model} and $H_0$, $T^+$ stochastically dominates $T$ in such a way that
\begin{equation} 
\label{eq:rosenbaum_bound}
\Pr(T \geq t \mid \mathcal{F},\mathcal{V}) \leq \Pr(T^+ \geq t \mid \mathcal{F},\mathcal{V}) \quad \quad \forall \bm{w} \in [-1,1]^I
\end{equation} 
\citep{rosenbaum1987, rosenbaum2002}. In \eqref{eq:rosenbaum_bound}, the upper bound is sharp, attained at $\bm{w} = \bm{w}^+$, and cannot be improved without further information about $\bm{u}$. 
In principle, this upper bound is examined by considering several values of $\Gamma = \exp(\gamma) \geq 1$ to search for the (truncated) \textit{sensitivity value}, the smallest $\Gamma$ that produces a worst-case $p$-value above a significance level $\alpha$. This sensitivity value then serves as an indicator of the robustness of the analysis to unmeasured confounding.

\subsection{The Power of a Sensitivity Analysis and Design Sensitivity}
\label{sec:review_design_sensi}

Suppose there is a treatment effect, meaning that $H_0$ is false. Also, assume that matching for $\bm{x}_{ij}$ successfully eliminated bias so that $\Pr(\bm{V}=\bm{v} \mid \mathcal{F},\mathcal{V}) = 1/2^I$ for each $\bm{v} \in \mathcal{V}$. If we were in this situation with an effect and no unmeasured bias from $\bm{u}$, called a \textit{favorable situation}, we could not be certain of this from the observed data, and the best we could hope to say is that the results are insensitive to moderate biases $\Gamma \geq 1$. The power of an $\alpha$ level sensitivity analysis is the probability that the worst-case $p$-value $\Pr(T^+ \geq t \mid \mathcal{F},\mathcal{V})$ will be less than or equal to $\alpha$, computed in a given favorable situation \citep{rosenbaum2004}. 
For a given model for generating $(\mathcal{F}, \mathcal{V})$ with an effect and no bias from $\bm{u}$, there typically exists a value $\tilde{\Gamma}$ called the design sensitivity such that, as $I \rightarrow \infty$, the power of a sensitivity analysis tends to $1$ if the analysis is performed with $\Gamma < \tilde{\Gamma}$ and to $0$ if performed with $\Gamma > \tilde{\Gamma}$. 
That is, in infinitely large sample sizes, the test can distinguish the model from biases smaller than $\tilde{\Gamma}$ but not from biases larger than $\tilde{\Gamma}$. 
See \cite{rosenbaum2004} and \cite{rosenbaum2010} for detailed discussions on design sensitivity, and see \citet{zhao2019} for insights into the relationship between sensitivity value and design sensitivity.


\section{Method}
\label{sec:method}

\subsection{Test of No Treatment Effect up to a Specific Time $\tau$}
\label{sec:time_specific_test}

We first consider testing the null hypothesis of no effect for any unit up to a given time point $\tau > 0$,
\begin{equation*}
H_0(\tau): \min(s_{0ij},\tau) = \min(s_{1ij}, \tau) \quad \text{for any} \ ij.
\end{equation*}

We define $q_{ij}(\tau)$ to be a time-specific score for unit $ij$ that is fixed under this null hypothesis by conditioning on $\mathcal{F}$. 
Although many choices are possible, we adopt the pseudo-observation approach of \citet{andersen2003}.
Let $\hat{K}(\tau)$ denote the Kaplan–Meier estimator based on all $2I$ units. The pseudo-observation of $\hat{K}(\tau)$ for unit $ij$ is defined as
$q_{ij}(\tau) = 2I\, \hat{K}(\tau) - (2I - 1)\, \hat{K}_{-ij}(\tau)$,
where $\hat{K}_{-ij}(\tau)$ is the Kaplan–Meier estimator based on $2I-1$ units $i^\prime j^\prime \neq ij$.
While $\hat{K}(\tau)$ can be replaced by an alternative estimator, we base $q_{ij}(\tau)$ on the Kaplan–Meier estimator throughout. 
See \citet{andersen2010} for an overview of the pseudo-observation approach, and \citet{kjaer2016} and \citet{andersen2017} for applied examples in causal survival analysis.
Intuitively, the time-specific score $q_{ij}(\tau)$ can be viewed as the contribution of unit $ij$ to the sample survival at time~$\tau$. 
Note that although $H_0(\tau)$ is only partially sharp, conditioning $\mathcal{F}$ still ensures that $q_{ij}(\tau)$ is fixed under this null hypothesis.
To see a concrete example, let $\tau>1$ and consider $I=1$ pair of two uncensored units. Suppose unit $11$ exhibits survival time $s_{111} = \tau+5$ if treated or survival time $s_{011} = \tau+1$ under control and unit $12$ exhibits $s_{112} = s_{012} = \tau-1$ whether treated or control. 
Then, $H_0(\tau)$ is true for this pair as $\min(s_{011},\tau) = \min(s_{111},\tau) = \tau$ and $\min(s_{012},\tau) = \min(s_{112},\tau) = \tau-1$, but the sharp null $H_0$ is false as $s_{111} \neq s_{011}$. 
Still, the scores $q_{11}(\tau)$ and $q_{12}(\tau)$ are fixed at $1$ and $0$, regardless of the value of $V_1 = Z_{11}-Z_{12}$, since neither $\hat{S}(\tau)$ nor the leave-one-out estimates $\hat{S}_{-11}(\tau)$ and $\hat{S}_{-12}(\tau)$ are affected by the discrepancy between $s_{111} = \tau+5$ and $s_{011} = \tau+1$.

In parallel to the notation in Section~\ref{sec:notation}, let $T_\tau= \sum_{i=1}^{I} d_i(\tau)V_i$, where $d_i(\tau) = q_{i1}(\tau) - q_{i2}(\tau)$, and $t_\tau$ be its observed value. 
Write $w_{\tau,i}^+ = \text{sgn}(d_i(\tau))$ and $\bm{w}^+_\tau = (w_{\tau,1}^+, \dots, w_{\tau,I}^+)^T$. 
Moreover, let $T^+_\tau$ denote the distribution of $T_\tau$ when $\bm{w}$ equals $\bm{w}^+_\tau$. 
Then, under \eqref{eq:rosenbaum_model} and $H_0(\tau)$, the test statistic $T_{\tau}$ satisfies
\begin{equation} \label{eq:time_specific_bound}
\Pr(T_\tau \geq t_\tau \mid \mathcal{F},\mathcal{V}) \leq \Pr(T^+_\tau \geq t_\tau \mid \mathcal{F},\mathcal{V}) \quad \forall \bm{w} \in [-,1,1]^I,
\end{equation}
and, as $I \rightarrow \infty$, the upper bound in \eqref{eq:time_specific_bound} (i.e., the worst-case $p$-value) is approximated by $\Pr(T^+_\tau \geq t_\tau |\mathcal{F},\mathcal{V}) \overset{\cdot}{=} 1 - \Phi\left((t - \mu_{\tau}^+)/\sigma_{\tau}^{+}\right),$ where $\mu_{\tau}^+$ and $(\sigma_{\tau}^{+})^2$ are the conditional mean and variance of $T^+_\tau$ given $(\mathcal{F},\mathcal{V})$, and $\Phi(\cdot)$ is the standard normal cumulative distribution. 
The asymptotic equivalence $\overset{\cdot}{=}$ holds if ${ \displaystyle \max_{1 \leq i \leq I} } d_i^2(\tau) / \sum_{i=1}^{I} d_i^2(\tau) \rightarrow 0$ as $I \rightarrow \infty$. 
The mean and variance are easily seen to be 
$\mu_{\tau}^+ = [(\Gamma-1)/(1+\Gamma)] \sum_{i=1}^{I} \lvert d_i(\tau) \rvert$ and $(\sigma_{\tau}^{+})^2 = [4\Gamma/(1+\Gamma)^2] \sum_{i=1}^I d_i^2(\tau)$.
When $\Gamma = 1$, the approximation yields an asymptotically exact randomization $p$-value, whereas for $\Gamma > 1$, it provides an asymptotically exact worst-case $p$-value.
Exact inferences can also be done by considering all $2^I$ possible values of $T^+_{\tau}$, when feasible. 


\subsection{Test of No Treatment Effect At All} \label{sec:overall_test}

We now return to the problem of testing the sharp null hypothesis $H_0$ of no effect at all. Although $H_0$ can be tested directly using the scores discussed in Section~\ref{sec:notation}, we choose to indirectly test it through the intersection of $L$ overlapping hypotheses,
\begin{equation*}
H_0^{*} = \cap_{l=1}^{L} H_0(\tau_{l}),
\end{equation*}
where $\tau_1, \dots, \tau_L$ are $L$ given times such that the numbers of events expected between them are sufficiently large.
The intersection null hypothesis $H_0^{*}$ is not identical to $H_0$, but includes it. When $\tau$ is sufficiently large or simply $\tau = \infty$, $H_0(\tau)$ should be the same as $H_0$. By definition, rejecting at least one of $H_0(\tau_{l}), l=1,\dots, L$ leads to the rejection of $H_0^*$, and consequently $H_0$.
As such, rejecting $H_0^*$ is more challenging if there is an effect. 
Existing tests typically target $H_0$ by aggregating survival differences between treated and control groups at each observed time point. 
However, if the differences vary across time points, then any significant difference at a specific time may be diluted when aggregated, resulting in reduced power.
Our proposed hypothesis, in contrast, examines survival differences across different time points and is thus more likely to capture any time-specific effects.  
In the remainder of this article, we refer to a test of $H_0$ as an overall test.

Assume that $H_0^*$ is true so that all associated time-specific scores $q_{ij}(\tau_l), i=1,\dots,I, j=1,2, l=1, \dots, L$ are fixed by conditioning on $\mathcal{F}$. Also, suppose for a moment that there is no concern about bias from $\bm{u}$ ($\Gamma=1$). Then, it is easy to see that the conditional covariance of $T_{\tau_k} = \sum_{i=1}^{I} d_i(\tau_k)V_i$ and $T_{\tau_l} = \sum_{i=1}^{I} d_i(\tau_l)V_i$ given $(\mathcal{F},\mathcal{V})$ is
$\sum_{i=1}^{I} d_i(\tau_k)d_i(\tau_l) = \sigma_{\tau_k\tau_l}$ for $1 \leq k, l \leq L$. 
Write $\sigma_{\tau_l}^2 = \sigma_{\tau_l\tau_l}$ for simplicity. Let $\rho_{\tau_k\tau_l} = \sigma_{\tau_k\tau_l}/(\sigma_{\tau_k}\sigma_{\tau_l})$ and $\bm{\rho}$ be the $L \times L$ correlation matrix containing $\rho_{\tau_k\tau_l}$. 
Accordingly, define 
$ M = \displaystyle \max_{1 \leq l \leq L} T_{\tau_l} /\sigma_{\tau_l} $ 
as the maximum of the $L$ associated standardized time-specific statistics.
With this notation, Theorem \ref{thm:thm1} states a multidimensional central limit theorem for the joint distribution of $T_{\tau_1}, \dots, T_{\tau_L}$, providing a way to conduct an asymptotically exact randomization test of $H_0^*$ using $M$ as the overall test statistic.

\begin{theorem} \label{thm:thm1}
Under the intersection null hypothesis $H_0^*$, if
\begin{equation} \label{cond:thm1} 
\max_{1 \leq l \leq L} \left\{\max_{1 \leq i \leq I} \frac{d_i^2(\tau_l)}{\sum_{i=1}^{I} d_i^2(\tau_l)}\right\} \rightarrow 0 \quad \text{ as }\> I \rightarrow \infty,
\end{equation}
then, in the absence of bias from $\bm{u}$ ($\Gamma=1$), the randomization $p$-value at the observed value $m$ of the overall test statistic $M$ is approximated by 
$\Pr(M \geq m \mid \mathcal{F}, \mathcal{V}) \overset{\cdot}{=} 1 - \bm{\Phi}_{\bm{\rho}}(m, \dots, m)$,
where $\bm{\Phi}_{\bm{\rho}}$ is the $L$-dimensional standard normal cumulative distribution function with correction matrix $\bm{\rho}$.
\end{theorem}
Unlike $p$-value combination methods, this approach adjusts for multiple testing error through direct computation of the dependence structure $\bm{\rho}$. 
Our overall test using $M$ is analogous to those in \citet{rosenbaum2012testing} and \citet{lee2018powerful}, which have been successful in multiple testing problems.

Theorem~\ref{thm:thm1} can be further extended to an asymptotically exact sensitivity analysis for a specific value of $\Gamma > 1$ under \eqref{eq:rosenbaum_model}. 
Recall that $T^+_{\tau_l}$ denotes the distribution of $T_{\tau_l}$ when $\bm{w} = \bm{w}^+_{\tau_l}$. 
Under our assumptions that $H_0^*$ is true, under \eqref{eq:rosenbaum_model}, it is easy to check that
\begin{equation} 
\label{eq:overall_bound}
\begin{split}
\Pr(M \geq m \mid \mathcal{F},\mathcal{V}) \leq \Pr \left( \max_{1 \leq l \leq L} T^+_{\tau_l}/\sigma_{\tau_l} \geq m \mid \mathcal{F},\mathcal{V} \right) \quad \forall \bm{w} \in [-1,1]^{I}.
\end{split}
\end{equation}
Without censoring, the signs of nonzero values among $d_i(\tau_1), \dots, d_i(\tau_L)$ are concordant for each pair $i$, 
so $\bm{w}_{\tau_1}^+ = \dots = \bm{w}_{\tau_L}^+$ (with the value of $\bm{w}_{\tau_l, i}^+$ for any tied pair $i$ with $d_i(\tau_l)=0$ reset to meet these equalities). Therefore, the worst-case $p$ value is achieved when $\bm{w} = \bm{w}_{\tau_1}^+ = \dots = \bm{w}_{\tau_L}^+$, meaning that the inequality \eqref{eq:overall_bound} is sharp. 
When observations are censored, however, it may not be true that $\bm{w}_{\tau_1}^+ = \dots = \bm{w}_{\tau_L}^+$. Consider the case where five pairs have observed times $\bm{Y} =(8.3,1.8,4.8,9.8,4.5,11.4,5.8,9.4,5.9,1.3)^T$ and event indicators $\bm{\Delta} = (1,1,1,1,1,0,0,1,1,1)^T$.
In this example, $\bm{V} = (1,1,1,1,1)^T$ produces $d_5(1.3) = -1$ and $d_5(5.9) = 1/5$ as the censored unit $41$ with $Y_{41} = 5.8$ is excluded from the ``at risk'' set before time $\tau = 5.9$.
With censoring, the upper bound in \eqref{eq:overall_bound} may be somewhat conservative but is readily solved by a normal approximation, as stated in Theorem~\ref{thm:thm2}. To formulate the theorem, we first introduce the following notation. Under our assumptions, it can easily be derived that the conditional covariance of $T_{\tau_k}^+$ and $T_{\tau_l}^+$ given $(\mathcal{F},\mathcal{V})$ is $\{4\Gamma/(1+\Gamma)^2\} \sum_{i=1}^{I} |d_i(\tau_k)d_i(\tau_l)| = \sigma_{\tau_k\tau_l}^+$ for $1 \leq k,l \leq L$. 
Write $\rho_{\tau_k\tau_l}^+ = \sigma_{\tau_k\tau_l}^+/(\sigma_{\tau_k}^+ \sigma_{\tau_l}^+)$ and $\bm{\rho}^+$ for the $L \times L$ correlation matrix containing $\rho_{\tau_k\tau_l}^+$.
Notice that $\sigma_{\tau_l \tau_l}^+$ is the same as $(\sigma_{\tau_l}^{+})^2$ defined in Section~\ref{sec:time_specific_test}, while, for $k \neq l$, $\sigma_{\tau_k\tau_l}^+$ does not necessarily equal $\sigma_{\tau_k\tau_l}$.

\begin{theorem} 
\label{thm:thm2}
Under the intersection null hypothesis $H_0^*$ and \eqref{eq:rosenbaum_model}, if the condition \eqref{cond:thm1} in Theorem~\ref{thm:thm1} holds, then, as $I \to \infty$, the worst-case $p$-value for a specific $\Gamma > 1$ is approximated by
$\Pr \left( {\displaystyle \max_{1 \leq l \leq L} } T^+_{\tau_l}/\sigma_{\tau_l} \geq m \mid \mathcal{F},\mathcal{V} \right) \overset{\cdot}{=} 1 - \bm{\Phi}_{\bm{\rho}^+}
\left( (m\sigma_{\tau_1}-\mu_{\tau_1}^+)/\sigma_{\tau_1}^+, \dots, (m\sigma_{\tau_L}-\mu_{\tau_L}^+)/\sigma_{\tau_L}^+ \right)$,
where $\bm{\Phi}_{\bm{\rho}^+}$ is the $L$-dimensional standard normal cumulative distribution function with correction matrix $\bm{\rho}^+$.
\end{theorem}

The approximations in Theorem~\ref{thm:thm1} and Theorem~\ref{thm:thm2} are adequate when the distributions $\bm{\Phi}_{\bm{\rho}}$ and $\bm{\Phi}_{\bm{\rho}^+}$ are not degenerate. This condition may easily hold when the selected time points $\tau_1, \dots, \tau_L$ are apart enough to limit high correlations between $T_{\tau_1}, \dots, T_{\tau_L}$.

\subsection{Design Sensitivities under Random Censoring}
\label{sec:design_sensitivity}

In this section, we derive design sensitivity formulas for the proposed tests in Section~\ref{sec:time_specific_test} and Section~\ref{sec:overall_test} to evaluate their asymptotic power in sensitivity analyses. 
Throughout this section, we assume that the realized survival and censoring times are independent, that is $S_{ij} \indep c_{ij}$. This assumption is commonly referred to as random censoring or non-informative censoring. 
As a first step, for a fixed $\tau$, we investigate the large-sample behavior of $q_{ij}(\tau), i = 1, \dots, I, j = 1, 2$  under random censoring.
Suppose there is a single cause-of-event, or equivalently, the analysis pertains to all-cause survival (i.e., no competing risks). Then, each $q_{ij}(\tau)$ is represented as
\begin{equation}
\label{eq:graw_represent}
K(\tau) - \dot{\psi}(Y_{ij}, \Delta_{ij}; \tau) + o_P(1),
\end{equation}
where $K(\tau) = \Pr(S_{ij} \geq \tau)$ and
$\dot{\psi}(\cdot; \tau)$ denotes the first-order influence function of the Aalen–Johansen functional \citep{aalen1978}.
Therefore, as $I \to \infty$, $q_{ij}(\tau), i = 1, \dots, I, j = 1, 2$, are approximately independent and identically distributed (i.i.d.). Consequently, as $I \to \infty$, the differences $d_i(\tau) = -\dot{\psi}(Y_{i1}, \Delta_{i1}; \tau) + \dot{\psi}(Y_{i2}, \Delta_{i2}; \tau) + o_P(1), i=1,\dots, I$ are also approximately i.i.d.
See \citet{graw2009} for a proof of the representation~\eqref{eq:graw_represent} and see \citep{andersen2010} for a discussion on large-sample properties of pseudo-observations.

Consider first whether $T_\tau$ would lead to rejection of $H_0(\tau)$ in a sensitivity analysis using the model \eqref{eq:rosenbaum_model}. By Theorem~\ref{thm:thm1}, as $I \rightarrow \infty$, the worst-case $p$-value is (asymptotically) less than or equal to the significance level $\alpha$ if 
\begin{equation} 
\label{eq:time_specific_rejection}
\Phi^{-1}(1-\alpha) \leq \frac{ \frac{\sum_{i=1}^{I} d_i(\tau)V_i/I}{\sqrt{\sum_{i=1}^I d_i^2(\tau)/I}} - \frac{\Gamma-1}{1+\Gamma} \frac{\sum_{i=1}^{I} |d_i(\tau)|/I}{\sqrt{\sum_{i=1}^I d_i^2(\tau)/I}} }{ \sqrt{\frac{4\Gamma}{(1+\Gamma)^2}/I} } = D_{\tau, I}.
\end{equation}
The event \eqref{eq:time_specific_rejection} means that $H_0(\tau)$ is rejected for all possible allocations of $\bm{w} \in [-1,1]^I$. The power of this $\alpha$ level sensitivity analysis is defined as the probability $\Pr(D_{\tau, I} \geq \Phi^{-1}(1-\alpha))$ under a given favorable situation with an effect and no bias from $\bm{u}$.
At $\Gamma=1$, this is just the power of a randomization test. As noted in Section~\ref{sec:review_design_sensi}, the design sensitivity of $T_{\tau}$ is the limiting sensitivity, $\tilde{\Gamma}_\tau$, such that, as $I \rightarrow \infty$, the power $\Pr(D_{\tau, I} \geq \Phi^{-1}(1-\alpha))$ tends to $1$ if the analysis if performed with $\Gamma < \tilde{\Gamma}_\tau$ and to $0$ if performed with $\Gamma > \tilde{\Gamma}_\tau$. Theorem~\ref{thm:thm3} states the detailed formula for $\tilde{\Gamma}_\tau$.

\begin{theorem}
\label{thm:thm3}
Consider a particular data-generating model for $(\mathcal{F}, \mathcal{V})$ with an effect and no bias from $\bm{u}$. Under the representation \eqref{eq:graw_represent}, if $\dot{\psi}(\cdot;\tau)$ is bounded, then the design sensitivity of the time-specific test statistic $T_{\tau}$ is 
\[
\tilde{\Gamma}_\tau = \frac{E( \lvert d_i(\tau) \rvert ) + E(d_i(\tau)V_i)}{E( \lvert d_i(\tau) \rvert ) - E(d_i(\tau)V_i)}.
\]
That is, for all $\alpha \in (0,1)$, as $I \rightarrow \infty$, the power $\Pr(D_{\tau, I} \geq \Phi^{-1}(1-\alpha)) \rightarrow 1$ if $\Gamma < \tilde{\Gamma}_\tau$ and $\Pr(D_{\tau, I} \geq \Phi^{-1}(1-\alpha)) \rightarrow 0$ if $\Gamma > \tilde{\Gamma}_\tau$.
\end{theorem}

Next, we examine the limiting sensitivity of $M$ to unmeasured bias. By Theorem~\ref{thm:thm2}, as $I \rightarrow \infty$, $H_0^*$ would be rejected for all possible allocations of $\bm{w} \in [-,1,1]^I$ if
\begin{equation}
\label{eq:overall_rejection}
1 - \alpha \leq \bm{\Phi}_{\bm{\rho}^+}
\left( \frac{ {\displaystyle \max_{1 \leq l \leq L}} \frac{\sum_{i=1}^I d_i(\tau_l)V_i/I}{\sqrt{\sum_{i=1}^{I} d_i^2(\tau_l)/I}} - \frac{\Gamma-1}{1+\Gamma} \frac{\sum_{i=1}^{I} |d_i(\tau_l)|/I}{\sqrt{\sum_{i=1}^{I} d_i^2(\tau_l)/I}} }{ \sqrt{\frac{4\Gamma}{(1+\Gamma)^2}/I} }, l=1,\dots,L \right) = \Lambda_{I}.
\end{equation}
In parallel, the power of this $\alpha$ level sensitivity analysis is the probability of the event \eqref{eq:overall_rejection} under a given favorable situation. But, in this case, for $\Gamma=1$, the probability $\Pr(\Lambda_I \geq 1-\alpha)$ is not necessarily the same as the power of a randomization test, since the inequality \eqref{eq:overall_bound} may not be equality at $\Gamma=1$ when the censoring occurred. We provide the formula for the design sensitivity of $M$ in Theorem~\ref{thm:thm4}.

\begin{theorem} 
\label{thm:thm4}
Consider a particular data-generating model for $(\mathcal{F}, \mathcal{V})$ with an effect and no bias from $\bm{u}$. Under the representation \eqref{eq:graw_represent}, if $\dot{\psi}(\cdot;\tau_l)$ is bounded for each $1 \leq l \leq L$, then the design sensitivity of the overall test statistic $M$ is 
\[
\tilde{\Gamma} = 
\frac{ {\displaystyle \max_{1 \leq l \leq L}} \frac{E(|d_i(\tau_l)|)}{\sqrt{E(d_i^2(\tau_l))}} + {\displaystyle \max_{1 \leq l \leq L}} \frac{E(d_i(\tau_l)V_i)}{\sqrt{E(d_i^2(\tau_l))}} }{ {\displaystyle \max_{1 \leq l \leq L}} \frac{E(|d_i(\tau_l)|)}{\sqrt{E(d_i^2(\tau_l))}} - {\displaystyle \max_{1 \leq l \leq L}}  \frac{E(d_i(\tau_l)V_i)}{\sqrt{E(d_i^2(\tau_l))}} }.
\]
That is, for all $\alpha \in (0,1)$, as $I \to \infty$, the power $\Pr(\Lambda_I \geq 1-\alpha) \to 1$ if $\Gamma < \tilde{\Gamma}$ and $\Pr(\Lambda_I \geq 1-\alpha) \to 0$ if $\Gamma > \tilde{\Gamma}$.
\end{theorem}

\subsection{Identifying the Effect Duration via Closed Testing} \label{sec:closed_testing}

When the overall null hypothesis $H_0$ is rejected through $H_0^{*}$, it may be of interest to identify the duration of the effect. 
Here, we formally determine the time points with an effect using the closed testing procedure of \citet{marcus1976closed}. 
Consider the intersection hypothesis $H_0^{A} = \cap_{\tau_l \in A} H_0(\tau_l)$, where $A$ is a subset of $\{\tau_1, \ldots, \tau_L\}$. Denote $\mathcal{I}_A \subset \{\tau_1, \ldots, \tau_L\}$ as any subset containing $A$ and $\mathcal{C}_A$ as the collection of all possible subsets $\mathcal{I}_A$. 
For example, given the set $\{\tau_1, \tau_2, \tau_3\}$ and $A = \{\tau_3\}$, $\mathcal{I}_{\{\tau_3\}}$ can be any element of the collection $\mathcal{C}_{\{\tau_3\}} = \left\{ \{\tau_3\}, \{\tau_1, \tau_3\}, \{\tau_2, \tau_3\}, \{\tau_1, \tau_2, \tau_3\} \right\}$. 
The closed testing procedure is simple, yet effectively controls the family-wise type 1 error rate. 
For each $A$, we reject $H_0^{A}$ at a significance level of $\alpha$ if all possible intersection hypotheses $H_0^{\mathcal{I}_A} = \cap_{\tau_l \in \mathcal{I}_A} H_0(\tau_l), \mathcal{I}_A \in \mathcal{C}_A$ involving $H_0^{A}$ are rejected at the same level. 
To be more precise, if we let $p(\mathcal{I}_{A})$ denote the $p$-value obtained in the test of $H_0^{\mathcal{I}_A}$, then $H_0^{A}$ is rejected only if the maximum $p$-value across all $\mathcal{I}_A \in \mathcal{C}_A$ is less than or equal to $\alpha$, i.e., $\displaystyle \max_{\mathcal{I}_A \in \mathcal{C}_A} p(\mathcal{I}_A) \leq \alpha$.
Computationally, to test $H_0^{\mathcal{I}_{A}}$, we derive a new test statistic $M^\prime = \displaystyle \max_{\tau_l \in \mathcal{I}_A} T_{\tau_l}/\sigma_{\tau_l}$ and obtain the $p$-value $p(\mathcal{I}_A)$ using the sub-correlation matrix $\bm{\rho}^\prime$ associated with the set $\mathcal{I}_A$. Since $\mathcal{I}_A \subset \{\tau_1, \ldots, \tau_L\}$, the multiple testing burden for testing $H_0^{\mathcal{I}_A}$ is less severe than for $H_0$. 
Although not guaranteed, in the case of $M^\prime = M$,  $p(\mathcal{I}_A) \leq \alpha$ when $p(\{\tau_1, \ldots, \tau_L\}) \leq \alpha$. The closed testing procedure can also be adapted for settings where $\Gamma > 1$. We will demonstrate in Section~\ref{sec:application} how the procedure can be implemented.

\section{Simulation}
\label{sec:simulation}

\subsection{Empirical Size and Power} \label{sec:size_power}


We perform a Monte Carlo simulation to evaluate the size and power of the proposed tests in Section~\ref{sec:time_specific_test} and Section~\ref{sec:overall_test}. 
For this simulation, we assume no unmeasured confounding ($\Gamma = 1$). 
To generate paired censored outcomes, we consider the hazard models $h(\tau|x_i, z) = \lambda \exp(x_i + \eta(\tau,z))$ for $s_{zij}$, $z=0,1$ and $\tilde{h}(\tau|x_i) = (\lambda/b) \exp(x_i)$ for $C_{ij}$, where $x_i \sim N(0,1)$ is the common variate for each pair, $\eta(\cdot)$ is an arbitrary function, $\lambda=0.2$, and $b>1$ is a constant for adjusting the censoring rate. We set the administrative censoring time to 5 to ensure that no observed time exceeds this limit. 

\begin{figure}[H] 

    \centering
    \begin{subfigure}[b]{0.19\textwidth}
        \includegraphics[width=\textwidth]{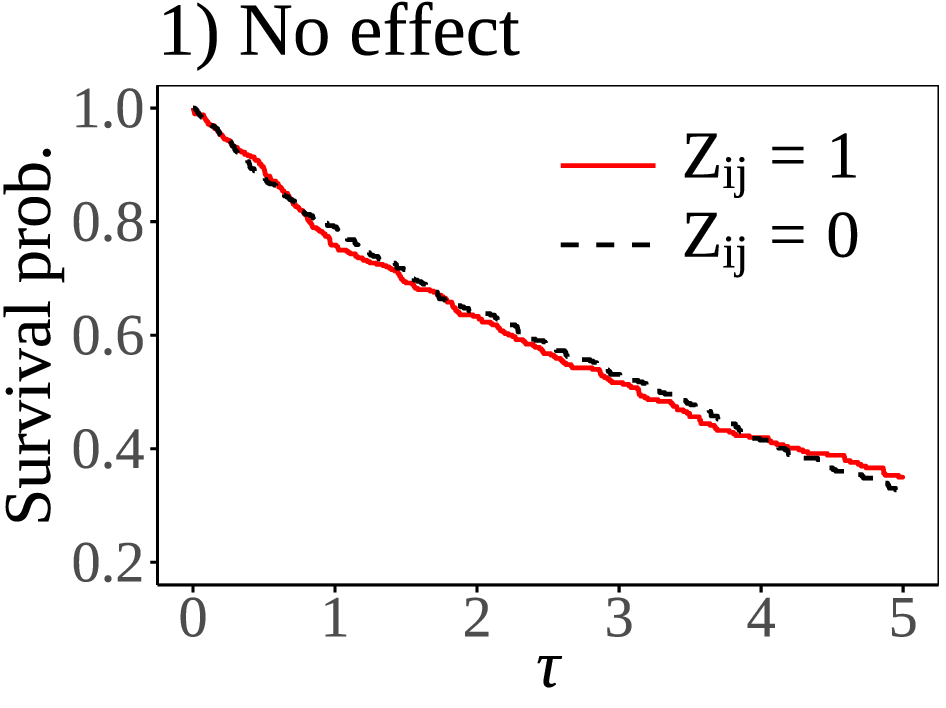}
    \end{subfigure}
    \begin{subfigure}[b]{0.19\textwidth}
        \includegraphics[width=\textwidth]{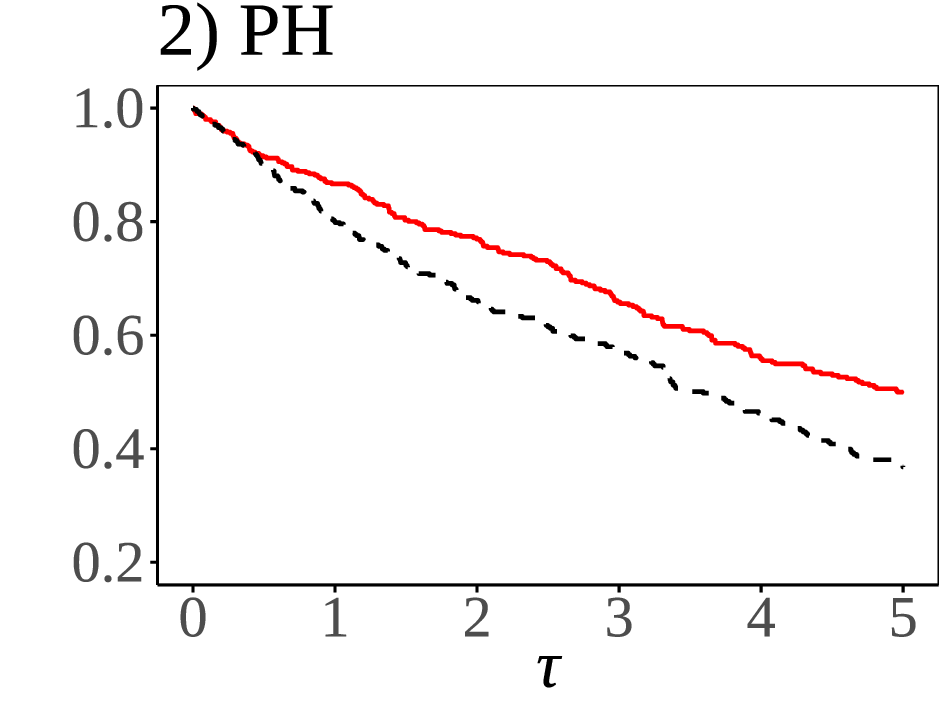}
    \end{subfigure}
    \begin{subfigure}[b]{0.19\textwidth}
        \includegraphics[width=\textwidth]{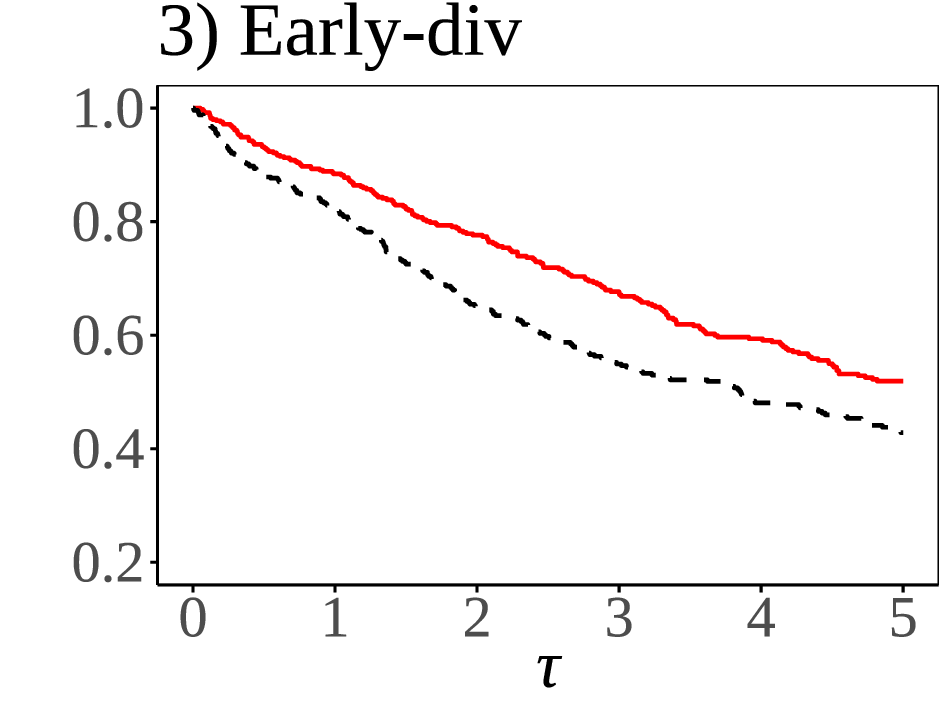}
    \end{subfigure}
    \begin{subfigure}[b]{0.19\textwidth}
        \includegraphics[width=\textwidth]{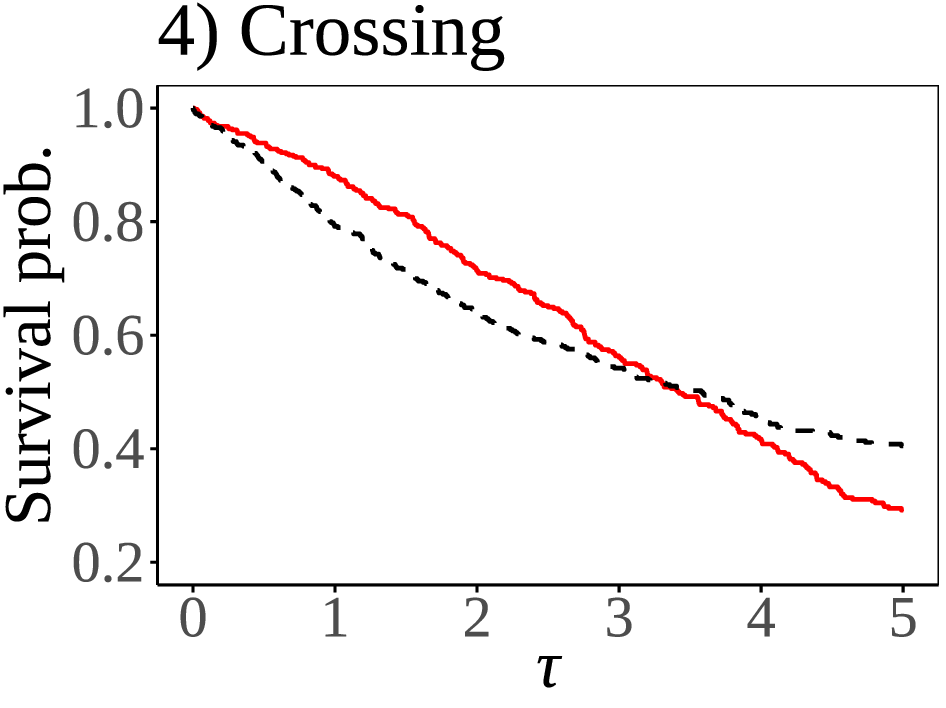}
    \end{subfigure}
    \begin{subfigure}[b]{0.19\textwidth}
        \includegraphics[width=\textwidth]{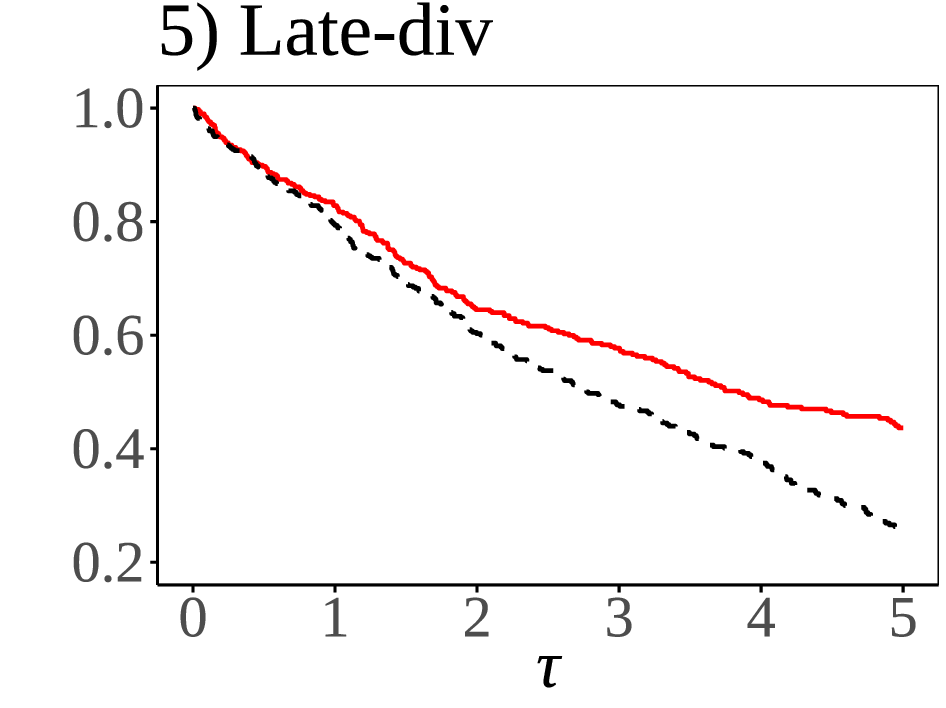}
    \end{subfigure}

    \captionsetup{width=0.9\textwidth}
    \caption{Kaplan-Meier survival curves for the example datasets of $I=500$ pairs simulated under the respective scenarios. }
    \label{fig:fig1}
    
\end{figure}

We generate $2000$ replications consisting of $I=500$ independent pairs of censored survival times for each of the following five scenarios: 1)  $\eta(\tau,z) = 0$; 2) $\eta(\tau, z) = -0.4z$; 3) $\eta(\tau,z) = (0.1\tau-0.5)z$; 4) $\eta(\tau,z) = (0.3\tau-0.6)z$; 5) $\eta(\tau,z) = (0.15-0.14z)\tau$. The first scenario indicates that no effect throughout follow-up (referred to as no effect). The second scenario corresponds to a proportional hazards model (referred to as PH). The third scenario (referred to as Early-div) and the fourth scenario (referred to as Crossing) both lead the survival curves for $z=0,1$ to diverge early on in favor of $z=1$, but the latter allows the curves to cross before the administrative censoring. In contrast, the fifth scenario leads the survival curves for $z=0,1$ to exhibit late divergence in favor of $z=1$ (referred to as late-div). The specifics of the data-generating processes follow \citep{austin2012}. For each of these five scenarios, $b$ is chosen separately to achieve moderate (approximately $25\%$) non-administrative censoring. In each pair, one is randomly labeled $Z_{ij}=1$ and the other is labeled $Z_{ij}=0$, with the realized $S_{ij}$ set equal $s_{1ij}$ or $s_{0ij}$ accordingly. Figure \ref{fig:fig1} shows the Kaplan-Meier survival curves for the example datasets of $I=500$ pairs simulated under the respective scenarios.

For illustration, we test $H_0(\tau)$, $\tau=1,2,3,4,5$ and $H_0$ through $H_0^{*} = \cap_{\tau=1}^{5} H_0(\tau)$. For comparison, we also consider the paired Prentice-Wilcoxon statistic \citep{obrien1987}, denoted by $PPW$, to test $H_0$. For ease of computation, we apply normal approximations to obtain randomization $p$-values. 

Table~\ref{tab:sizepower} summarizes the empirical rejection rates at $\alpha=0.05$.
Under the No effect scenario, all tests maintain their empirical size close to the nominal level $\alpha=0.05$. In the PH scenario, $T_{\tau+1}$ exhibits a higher power relative to $T_{1} \dots, T_{\tau}$ despite the fact that the scenario assumes a constant effect over time. This is not surprising since the number of informative pairs with $d_i(\tau) \neq 0$ increases as time elapses, which also explains why $T_1$ has slightly lower power than $T_2$ in the early-div scenario. Other than this, the time-specific tests concordantly reflect the effect trajectories assumed in the respective scenarios.
In overall tests, the results of $PPW$ agree with those in \citep{woolson1992, dallas2000, sun1996, lee2003}, suffering from power loss in the Late-div and Crossing scenarios. Overall, $M$ outperforms $PPW$, although it has slightly lower power in the Early-div scenario. In particular, the power of $M$ is considerably higher compared to $PPW$ in the Crossing and Late-div scenarios.

\begin{table}[h] 
\centering \renewcommand{\tabcolsep}{6pt} \renewcommand{\arraystretch}{0.8}
\begin{threeparttable}
\caption{Empirical rejection rates at $\alpha=0.05$ computed based on $2000$ replications of $I = 500$ pairs}
\label{tab:sizepower}
\begin{tabular}{llcccccccc}
\hline \hline
\multicolumn{2}{l}{} & \multicolumn{5}{c}{$H_0(\tau)$} & & 
\multicolumn{2}{c}{$H_0$} \\ \cline{3-7} \cline{9-10}
Scenario & & $\tau=1$ & $\tau=2$ & $\tau=3$ & $\tau=4$ & $\tau=5$ & & $M$ & $PPW$ \\
\hline
1) No effect & & 0.052 & 0.047 & 0.052 & 0.055 & 0.050 & & 0.049 & 0.056 \\
2) PH        & & 0.782 & 0.939 & 0.963 & 0.979 & 0.982 & & 0.985 & 0.978 \\
3) Early-div & & 0.865 & 0.943 & 0.937 & 0.873 & 0.758 & & 0.954 & 0.957 \\
4) Crossing  & & 0.875 & 0.770 & 0.296 & 0.017 & 0.000 & & 0.798 & 0.375 \\
5) Late-div  & & 0.124 & 0.345 & 0.657 & 0.880 & 0.970 & & 0.927 & 0.610 \\
\hline
\end{tabular} 
\end{threeparttable}
\end{table}

\subsection{Design Sensitivities}
We now use the explicit formulas in Theorem~\ref{thm:thm3} and Theorem~\ref{thm:thm4} to asymptotically evaluate the power of sensitivity analyses using $T_\tau$ and $M$ via their design sensitivities. Since the formulas are valid under random censoring, in this simulation, we consider the alternative hazard model $\tilde{h}(\tau) = \lambda/b$ for $C_{ij}$, where $\lambda=0.2$ and $b>1$ is a constant for adjusting the censoring rate. Except for this, the settings are the same as those in Section~\ref{sec:size_power}. For each of the four alternative scenarios 2-5), $b$ is adjusted separately to achieve moderate (approximately $25\%$) non-administrative censoring. We calculate design sensitivities for $T_\tau$, $\tau=1,2,3,4,5$ and $M$ under each scenario by the following two steps: we first generate a single sample of $I=100,000$ independent pairs of censored survival times and then use the Monte Carlo method to approximate the expectations involved in formulas in Theorem~\ref{thm:thm3} and Theorem~\ref{thm:thm4}. 
The results are given in Table \ref{tab:ds}.
In the PH scenario, the design sensitivity of $T_\tau$ slightly increases as $\tau$ increases from $1$ to $5$. Under the three non-PH scenarios 3-5), the design sensitivities of the time-specific tests are relatively large at time points with strong effects, indicating greater robustness to unmeasured confounding. In the Crossing scenario, the adverse effect at $\tau = 4$ and $5$ leads the design sensitivities of $T_4$ and $T_5$ to be less than $1$.

\begin{table}[h]
\vspace{0.5cm}
\centering \renewcommand{\tabcolsep}{6pt} \renewcommand{\arraystretch}{0.8}
\begin{threeparttable}
\caption{Design sensitivities computed based on Monte Carlo samples of $I = 100,000$ pairs under random censoring}
\label{tab:ds}
\begin{tabular}{llccccccc}
\hline \hline
\multicolumn{2}{l}{} & \multicolumn{5}{c}{$\tilde{\Gamma}$} & & $\Gamma$ \\ 
\cline{3-7} \cline{9-9}
Scenario & & $\tau=1$ & $\tau=2$ & $\tau=3$ & $\tau=4$ & $\tau=5$ & & \\
\hline
2) PH        & & 1.491 & 1.530 & 1.549 & 1.560 & 1.568 & & 1.567 \\
3) Early-div & & 1.557 & 1.524 & 1.468 & 1.394 & 1.325 & & 1.465 \\
4) Crossing  & & 1.574 & 1.371 & 1.156 & $<$ 1 & $<$ 1 & & 1.398 \\
5) Late-div  & & 1.070 & 1.160 & 1.271 & 1.399 & 1.574 & & 1.524 \\
\hline
\end{tabular} 
\end{threeparttable}
\end{table}

\section{Application to the KLoSA Data} 
\label{sec:application}

We now apply our method to the matched pairs from the KLoSA cohort to evaluate how long social engagement at baseline in $2006$ continued to impact survival in the cohort. 
Since the KLoSA is a biennial survey, we consider the null hypotheses $H_0(\tau)$ at follow-up years $\tau = 2,4,6,8,10,12,14$ and $H_0^{*} = \cap_{\tau \in \{2,4,6,8,10,12,14\}} H(\tau)$ (to test $H_0$). In addition to $M$, we use two additional test statistics to test $H_0$: the paired Prentice-Wilcoxon statistic $PPW$ and the combined statistic $M+PPW$ which includes the standardized $PPW$ as a component of $M$.
As we noted in Section~\ref{sec:motive}, concerns about unmeasured confounding remain in this study, so we search for sensitivity values $\Gamma \geq 1$ that produce worst-case $p$-values above $\alpha=0.05$.
Table~\ref{tab:klosa_sensi} reports the worst-case $p$-values for various hypothesized $\Gamma$ values.

\begin{table}[H] 
\centering 
\renewcommand{\tabcolsep}{3pt} \renewcommand{\arraystretch}{0.8}
\caption{Worst-case $p$-values in sensitivity analyses for the 1,474 matched KLoSA pairs}
\label{tab:klosa_sensi}
\begin{threeparttable}
\begin{tabular}{llccccccccccc}
\hline \hline
\multicolumn{2}{l}{} & \multicolumn{7}{c}{$H_0(\tau)$} & & \multicolumn{3}{c}{$H_0$} \\ 
\cline{3-9} \cline{11-13}
Bias & & $\tau=2$ & $\tau=4$ & $\tau=6$ & $\tau=8$ & $\tau=10$ & $\tau=12$ & $\tau=14$ & & $M$ & $PPW$ & $M + PPW$\\
\hline
$\Gamma=1$    & & 0.001&	0.000&	0.000&	0.000&	0.005&	0.013&	0.037&&	0.000&	0.003 & 0.000 \\ 
$\Gamma=1.1$  & & 0.003&	0.000&	0.002&	0.000&	0.053&	0.128&	0.273&&	0.002&	0.039 & 0.002 \\ 
$\Gamma=1.15$ & & 0.005&	0.000&	0.004&	0.002&	0.120&	0.262&	0.480&&	0.009&	0.091 & 0.009 \\ 
$\Gamma=1.2$  & & 0.007&	0.001&	0.010&	0.005&	0.227&	0.438&	0.684&&	0.030&	0.179 & 0.031 \\ 
$\Gamma=1.25$ & & 0.011&	0.001&	0.020&	0.012&	0.366&	0.620&	0.838&&	0.082&	0.302 & 0.084 \\ 
$\Gamma=1.3$  & & 0.016&	0.002&	0.036&	0.027&	0.518&	0.773&	0.930&&	0.179&	0.445 & 0.182 \\ 
$\Gamma=1.35$ & & 0.021&	0.004&	0.061&	0.054&	0.663&	0.880&	0.974&&	0.321&	0.591 & 0.324 \\ 
$\Gamma=1.4$  & & 0.029&	0.007&	0.096&	0.094&	0.783&	0.943&	0.992&&	0.490&	0.720 & 0.493 \\ 
$\Gamma=1.45$ & & 0.038&	0.012&	0.142&	0.152&	0.871&	0.976&	0.998&&	0.656&	0.823 & 0.658 \\ 
$\Gamma=1.5$  & & 0.049 &	0.019&	0.199&	0.226&	0.929&	0.991&	0.999&&	0.792&	0.895 & 0.794 \\ 
$\Gamma=1.55$ & & 0.061&	0.028&	0.264&	0.314&	0.964&	0.997&	1.000&&	0.888&	0.943 & 0.888 \\ 
$\Gamma=1.6$  & & 0.075&	0.041&	0.337&	0.411&	0.983&	0.999&	1.000&& 0.945&	0.970 & 0.946 \\ 
$\Gamma=1.65$ & & 0.092&	0.057&	0.414&	0.510&	0.992&	1.000&	1.000&& 0.976&	0.986 & 0.976 \\ 
\hline
\end{tabular}
\end{threeparttable}
\end{table}

We find that, in the absence of unmeasured confounding ($\Gamma=1$), the $p$-values for all time-specific tests fall below $\alpha = 0.05$. As $\Gamma$ increases, we can see that the worst-case $p$-values for the time-specific tests vary across given time points. 
The treatment effects at $\tau=2, 4, 6, 8$ are quite strong even under larger values of $\Gamma$. 
For instance, we can reject $H_0(4)$ up to $\Gamma = 1.6$ and reject $H_0(8)$ up to $\Gamma=1.3$. 
However, the effects at later time points can be comparatively weak, and the $p$-values rapidly approach to 1 as $\Gamma$ increases. These effects can be explained if there is a small amount of unmeasured bias. This observed pattern is similar to the Early-div scenario in Section~\ref{sec:simulation}. 
In the Early-div scenario, there was a slight difference between $M$ and $PW$ in terms of power. 
However, in the sensitivity analyses, the inference from $M$ is more robust to unmeasured bias. 
Regarding the overall tests, $M$ leads to rejection of $H_0$ for $\Gamma \leq 1.2$ and $PW$ lead to rejection of $H_0$ for $\Gamma \leq 1.1$. 
We also find that our overall test using $M$ can be effectively integrated with $PPW$. 
The results of $M+PPW$ are nearly identical to those of $M$ alone, indicating that the addition of $PPW$ does not enhance performance. 
Surprisingly, there is a negligible contribution to multiple corrections by incorporating $PPW$.

As discussed in Section~\ref{sec:closed_testing}, we next apply the closed testing procedure in Section\ref{sec:closed_testing} to test each $H_0(\tau)$ more thorougly. We report the maximum $p$-values $\displaystyle \max_{\mathcal{I}_A \in \mathcal{C}_A} p(\mathcal{I}_A), A \in \{\{2\}, \{4\}, \{6\}, \{8\}, \{10\}, \{12\}, \{14\}\}$ for various hypothesized $\Gamma \geq 1$ values in Table~\ref{tab:klosa_closed_test}. 
In the absence of unmeasured bias ($\Gamma=1$), all hypotheses are rejected at $\alpha = 0.05$. This suggests that the effect of social engagement at baseline on survival in the cohort sustained through 2022, assuming no unmeasured confounding. Overall, after correction, the inferences become more conservative. For example, $H_0(4)$ was rejected up to $\Gamma = 1.6$, but after correction, it is now rejected up to $\Gamma = 1.2$. 
In summary, we conclude that social engagement at baseline in 2006 continued to prolong survival throughout follow-up, initially strong but diminishing over time.

\begin{table}[H] 
\centering 
\renewcommand{\tabcolsep}{6pt} 
\renewcommand{\arraystretch}{0.8}
\caption{Maximum $p$-value bounds $\displaystyle \max_{\mathcal{I}_A \in \mathcal{C}_A} p(\mathcal{I}_A)$ in closed testing for the 1,474 matched KLoSA pairs}
\label{tab:klosa_closed_test}
\begin{threeparttable}
\begin{tabular}{llccccccc}
\hline \hline
\multicolumn{2}{l}{} & \multicolumn{7}{c}{$H_0^{A}$} \\ 
\cline{3-9}
Bias & & $A=\{2\}$ & $A=\{4\}$ & $A=\{6\}$ & $A=\{8\}$ & $A=\{10\}$ & $A=\{12\}$ & $A=\{14\}$ \\
\hline
$\Gamma=1$    & & 0.003& 0.000&	0.001&	0.000&	0.011&	0.021&	0.037 \\
$\Gamma=1.1$  & & 0.045& 0.002&	0.016&	0.003&	0.121&	0.187&	0.273 \\
$\Gamma=1.15$ & & 0.120& 0.009&	0.051&	0.014&	0.262&	0.365&	0.480 \\
$\Gamma=1.2$  & & 0.252& 0.030&	0.129&	0.046&	0.451&	0.570&	0.684 \\
$\Gamma=1.25$ & & 0.430& 0.082&	0.260&	0.116&	0.645&	0.751&	0.838 \\
\hline
\end{tabular}
\end{threeparttable}
\end{table}

\section{Discussion}

In this article, we introduce a powerful randomization-based method for granularly evaluating treatment effects in paired cohort studies. Our contribution is manifold. 
First, our method provides a principled way to identify the onset and duration of a treatment effect. This strategy will be particularly useful for studies seeking to demonstrate a treatment's efficacy within a set time or to understand the temporal behavior of an effect. 
Second, we showed that the overall test adapts to a range of time-varying effects, complementing existing tests whose power may vary substantially with the underlying effect trajectory. Third, we incorporated sensitivity analysis procedures, extending the applicability of our method to observational studies where randomization may not hold.

We suggest three future research directions based on our findings. First, our method can be improved by adopting more effective time-specific scores than pseudo-observations. 
Particularly at early time points, where events rarely occur for most individuals, many scores are uninformative. 
If there is an alternative approach of computing scores during these early periods, any initial difference can be more easily detected. Second, the inequality~\eqref{eq:overall_bound} may not be sufficiently precise when $\Gamma>1$. 
A more computationally intensive strategy could be explored to improve its bound, as suggested by \citet{fogarty2016sensitivity}. Implementing such a method requires effectively integrating the correlation structure of the test statistics. 
Lastly, in our application, the time points were predetermined based on biennial data collection. 
A more interesting approach would involve developing a data-driven approach that adaptively selects the number and location of time points to analyze the effect. This could lead to more tailored and potentially insightful findings.

\section*{Data Availability}
The data used in this study was obtained on 23rd June 2024 from the Korea Employment Information Service (KEIS) website, publicly available at \url{https://survey.keis.or.kr/eng/index.jsp}.

\section*{Disclosure Statement}
The authors report there are no competing interests to declare.

\bibliographystyle{apalike} 
\bibliography{ref.bib}

\appendix
\renewcommand{\thesection}{Appendix~\Alph{section}}

\section{Technical Proofs} \label{sec:append_proof}

\subsection{Proof of Theorem 1}




Theorem \ref{thm1} follows from the multivariate central limit theorem by \citet[p.~164]{fabian1985}, taking their $n$ to be $I$ and their $\bm{X}_{nj}$ to be $( d_j(\tau_1)/\sigma_{\tau_1}, \dots, d_j(\tau_L)/\sigma_{\tau_L})^TV_j$. Define $\bm{e}_l$ to be the unit vector in $\mathbb{R}^L$ with its $l$th entry being $1$. By the argument in the central limit theorem, it is enough to verify that, for each $\bm{e}_l$, the triangular array $\bm{e}_l^T\bm{X}_{n1}, \dots, \bm{e}_l^T\bm{X}_{nn}$, $n \geq 1$ satisfy the Lindberg condition,
\begin{equation*}
\begin{split}
\lim_{n\to\infty} \sum_{j=1}^{n} E\left[ (\bm{e}_l^T\bm{X}_{nj})^2 I(|\bm{e}_l^T\bm{X}_{nj}| > \epsilon) | \mathcal{F}, \mathcal{V} \right] = 0 \ \text{for any} \ \epsilon > 0.
\end{split}
\end{equation*} 
For each $\tau_l \in \mathcal{T}$ and a given $\epsilon > 0$, 
\begin{equation*}
\begin{split}
&\sum_{j=1}^{n} E\left[ (\bm{e}_l^T\bm{X}_{nj})^2 I(|\bm{e}_l^T\bm{X}_{nj}| > \epsilon) | \mathcal{F}, \mathcal{V} \right] \\
&= \sum_{j=1}^{n} E\left[ \left( d_j(\tau_l)V_j/\sigma_{\tau_l} \right)^2 I\left( \left| d_j(\tau_l)V_j/\sigma_{\tau_l} \right| > \epsilon \right) \bigg| \mathcal{F}, \mathcal{V} \right] \\[0.5em]
&= \sum_{j=1}^{n} (d_j(\tau_l)/\sigma_{\tau_l})^2 E\left[V_j^2 I\left( d_j^2(\tau_l)V_j^2/\sigma_{\tau_l}^2 > \epsilon^2 \right) \bigg| \mathcal{F}, \mathcal{V} \right] \\[0.5em]
&\leq E\left[V_1^2 I\left( \max_{j \in \{1,\dots,n\}} \frac{d_j^2(\tau_l)}{\sum_{j=1}^{n} d_j^2(\tau_l)} V_1^2 > \epsilon^2 \right) \bigg| \mathcal{F}, \mathcal{V} \right] \\[1em]
&\leq E\left[
V_1^2 I\left( \max_{\tau_l \in \mathcal{T}} \left\{\max_{j \in \{1,\dots,n\}} \frac{d_i^2(\tau_l)}{\sum_{j=1}^{n} d_j^2(\tau_l)}\right\} V_1^2 > \epsilon^2 \right) \bigg| \mathcal{F}, \mathcal{V} \right],       
\end{split}
\end{equation*} 
where the third relation holds under randomized treatment assignments. Using the condition in Theorem~\ref{thm1}, the last line converges to $0$ as $n \rightarrow \infty$ by the dominated convergence theorem, proving Theorem~\ref{thm1}. Strictly speaking, as $n \rightarrow \infty$ in this proof, the notation should refer to sequences of $\mathcal{F}_n$ and $\mathcal{V}_n$, changing with $n$, but to avoid cumbersome notation, this was not indicated explicitly. $\square$

\subsection{Proof of Theorem 2}

The proof of Theorem \ref{thm2} is parallel to that of Theorem \ref{thm1}, taking $n$ to be $I$ and $\bm{X}_{nj}$ to be $((d_j(\tau_1)V_{\tau_1,j}^+ - \{(\Gamma-1)/(1+\Gamma)\}|d_j(\tau_1)|)/\sigma_{\tau_1}^+, \dots, (d_j(\tau_L)V_{\tau_L,j}^+ - \{(\Gamma-1)/(1+\Gamma)\}|d_j(\tau_L)|)/\sigma_{\tau_L}^+)^T$ where $V_{\tau_l,j}^+$ denotes the distribution of $V_j$ when $w_j = w_{\tau_l,j}^+$. $\square$

\subsection{Proof of Theorem 3}

Due to the expression~(\ref{grawapprox}), we can rewrite $d_i(\tau)V_i$ as $-( \dot{\psi}(Y_{i1},\delta_{i1},\tau) - \dot{\psi}(Y_{i2},\delta_{i2},\tau) ) V_i + o_P(1)$. As $I \to \infty$, using that $\dot{\psi}(\cdot;\tau)$ is bounded, the numerator of $D_{\tau, I}$ converges in probability to $\{E(d_i(\tau)V_i) - [(\Gamma-1)/(1+\Gamma)] E(|d_i(\tau)|)\}/\sqrt{E(d_i^2(\tau))}$ while the denominator of $D_{\tau, I}$ tends to $0$. The formula (\ref{design_sensitivity1}) is then obtained by rearranging the equation $\{E(d_i(\tau)V_i) - [(\Gamma-1)/(1+\Gamma)] E(|d_i(\tau)|)\}/\sqrt{E(d_i^2(\tau))} = 0$. $\square$

\subsection{Proof of Theorem 4}

Consider the limiting behavior of the minimum among
\begin{equation*}
\frac{ {\displaystyle \max_{\tau_l \in \mathcal{T}}} \frac{\sum_{i=1}^I d_i(\tau_l)V_i/I}{\sqrt{\sum_{i=1}^{I} d_i^2(\tau_l)/I}} - \frac{\Gamma-1}{1+\Gamma} \frac{\sum_{i=1}^{I} |d_i(\tau_l)|/I}{\sqrt{\sum_{i=1}^{I} d_i^2(\tau_l)/I}} }{ \sqrt{\frac{4\Gamma}{(1+\Gamma)^2}/I} } = D_{\tau_l,I}^*, \tau_l \in \mathcal{T},
\end{equation*}
as $I \to \infty$. By a similar argument to the proof of Theorem \ref{thm3}, as $I \to \infty$, the numerator of $\min_{\tau_l \in \mathcal{T}} D_{\tau_l, I}^*$ converges in probability to $\max_{\tau_l \in \mathcal{T}} E(d_i(\tau_l)V_i)/\sqrt{E(d_i^2(\tau_l))} - [(\Gamma-1)/(1+\Gamma)] \max_{\tau_l \in \mathcal{T}} E(|d_i(\tau_l)|)/\sqrt{E(d_i^2(\tau_l))}$ while the denominator of $\min_{\tau_l \in \mathcal{T}} D_{\tau_l, I}^*$ tends to $0$. Observing that 
\begin{equation*}
\Pr\left(\chi_l \leq \min_{\tau_l \in \mathcal{T}} D_{\tau_l, I}^*, l=1,\dots,L\right) \leq 
\Lambda_{\mathcal{T},I} \leq \Pr\left(\chi_1 \leq \min_{\tau_l \in \mathcal{T}} D_{\tau_l, I}^* \right),
\end{equation*}
$\Lambda_{\mathcal{T},I}$ tends to $1$ for $\Gamma < \tilde{\Gamma}_\mathcal{T}$ and to $0$ for $\Gamma > \tilde{\Gamma}_\mathcal{T}$ as $I \to \infty$. $\square$

\section{Details on the Matching} \label{sec:append_matching}


\begin{table}[H]
\centering \renewcommand{\tabcolsep}{4pt} \renewcommand{\arraystretch}{0.6}
\begin{threeparttable}
\caption{Confounder balance in 1474 matched treated-control pairs. We report the mean values of confounders and standardized mean differences (SMDs) before and after matching.}
\label{tab:balancetab}
\begin{tabular}{llcclcccl}
\hline \hline
\multicolumn{2}{l}{} & \multicolumn{3}{c}{Before Matching} & & \multicolumn{3}{c}{After Matching} \\ \cline{3-5} \cline{7-9}
Confounder & & Treated & Control & SMD & & Treated & Control & SMD \\
\hline
Age && 59.17& 62.30& -0.305&& 59.17& 59.18&	-0.001 \\
Age $\geq$ 65 && 0.315& 0.426& -0.237&&	0.315& 0.315& 0.000 \\
Male && 0.423&	0.442& -0.037&&	0.423& 0.423& 0.000 \\
Education &&&&&&&& \\
\hspace{0.5cm} $\leq$ Elementary school && 0.308& 0.519& -0.457&& 0.308& 0.310& -0.004 \\
\hspace{0.5cm} Middle school            && 0.134& 0.172& -0.111&& 0.134& 0.124& 0.030 \\
\hspace{0.5cm} High school              && 0.341& 0.241& 0.211&& 0.341& 0.386& -0.096 \\
\hspace{0.5cm} $\geq$ College           && 0.215& 0.068& 0.357&& 0.215& 0.179& 0.088 \\
\hspace{0.5cm} Missing                  && 0.002& 0.000& 0.040&& 0.002& 0.001& 0.030 \\
Urban Resident && 0.832& 0.794&	0.101&&	0.832& 0.827& 0.013 \\
Married && 0.830& 0.776& 0.144&& 0.830& 0.828& 0.004 \\
Income                  &&&&&&&& \\
\hspace{0.5cm} Low     && 0.200& 0.265& -0.163&& 0.200&	0.214& -0.034 \\
\hspace{0.5cm} Middle  && 0.246& 0.259& -0.029&& 0.246& 0.242& 0.011 \\
\hspace{0.5cm} High    && 0.305& 0.209& 0.209&& 0.305& 0.305& 0.000 \\
\hspace{0.5cm} Missing && 0.249& 0.267&	-0.043&& 0.249&	0.240& 0.020 \\
Economic Activity && 0.406&	0.359& 0.094&& 0.406& 0.409& -0.007 \\
Alcohol use && 0.428& 0.447& -0.038&& 0.428& 0.432&	-0.008 \\
Smoker && 0.244& 0.308&	-0.149&& 0.244&	0.254& -0.024 \\
Depression && 0.097& 0.130&	-0.112&& 0.097&	0.087& 0.034 \\
Self-reported health status &&&&&&&& \\
\hspace{0.5cm} Bad          && 0.177& 0.337&	-0.418&& 0.177&	0.189& -0.032 \\
\hspace{0.5cm} Moderate     && 0.460& 0.468& -0.017&& 0.460& 0.467&	-0.015 \\
\hspace{0.5cm} Good         && 0.363& 0.195& 0.349&& 0.363&	0.343& 0.041 \\
No. chronic diseases    &&&&&&&& \\
\hspace{0.5cm} 0        && 0.540& 0.510& 0.060&& 0.540& 0.547&	-0.015 \\
\hspace{0.5cm} 1        && 0.298& 0.291& 0.014&& 0.298& 0.286&	0.025 \\
\hspace{0.5cm} $\geq$ 2 && 0.162& 0.199& -0.099&& 0.162& 0.166&	-0.011 \\
\hline
\end{tabular} 
\end{threeparttable}
\end{table}



\end{document}